\documentclass{article}
\usepackage[preprint]{spconf}
\copyrightnotice{© 2024 IEEE.}
\toappear{To appear in {\it Proc.\ ICASSP2024, April 14-19, 2024, Seoul, Korea}}

\usepackage{spconf,amsmath,graphicx}
\usepackage{cite}
\usepackage{amsmath,amssymb,amsfonts}
\usepackage{algorithm}
\usepackage{algorithmicx}
\usepackage{algpseudocode}
\algdef{SE}[SUBALG]{Indent}{EndIndent}{}{\algorithmicend\ }%
\algtext*{Indent}
\algtext*{EndIndent}
\usepackage{lipsum}

\usepackage{threeparttable} % to use table notes
\usepackage[export]{adjustbox}
\usepackage[table]{xcolor}% http://ctan.org/pkg/xcolor
\usepackage{url}
\usepackage{bm}
\usepackage{courier}
\usepackage{bbm}
\usepackage{caption}
\usepackage{subfig}
\usepackage{braket}

\usepackage{array}
\usepackage{multirow}
\usepackage{graphicx}
\usepackage{tablefootnote}
\usepackage[T1]{fontenc}
\usepackage{booktabs}
\usepackage{multirow}
\usepackage{hyperref}

\usepackage{textcomp}
\usepackage{xcolor}

\title{Differentiable Quantum Architecture Search for Job Shop Scheduling Problem}
\name{Yize Sun$^{1,2}$ \qquad Jiarui Liu$^{1}$ \qquad Yunpu Ma$^{1,2}$ \qquad Volker Tresp$^{1,2}$}
\address{$^{1}$Ludwig-Maximilians-University Munich \qquad $^{2}$Siemens AG \\ Munich, Germany}

\usepackage{tikz}
\usepackage{textcomp}
\usepackage{hyperref}
\usepackage{lipsum}

\begin{document}
%\ninept
%
\maketitle
\begin{abstract}
  The Job shop scheduling problem (JSSP) plays a pivotal role in industrial applications,
  such as signal processing (SP) and steel manufacturing,
  involving sequencing machines and jobs to maximize scheduling
  efficiency. Before, JSSP was solved using manually defined circuits by variational quantum algorithm (VQA).
  Finding a good circuit architecture is task-specific and time-consuming.
  Differentiable quantum architecture search (DQAS) is a gradient-based framework
  that can automatically design circuits.
  However, DQAS is only tested on quantum approximate optimization algorithm (QAOA)
  and error mitigation tasks.
  Whether DQAS applies to JSSP based on a more flexible algorithm, such as
  variational quantum eigensolver (VQE), is still open for optimization problems.
  In this work, we redefine the operation pool and extend DQAS to a framework JSSP-DQAS
  by evaluating circuits to
  generate circuits for JSSP automatically.
  The experiments conclude that JSSP-DQAS can automatically find
  noise-resilient circuit architectures that perform much better than manually designed
  circuits. It helps to improve the efficiency of solving JSSP.
\end{abstract}
\section{Introduction}
\label{Introduction}
Quantum computing (QC) and quantum machine learning have rapidly developed in many areas \cite{biamonte2017quantum, schuld2015introduction, casana2020probabilistic, li2020quantum, jeswal2019recent, dong2008quantum, schuld2014quest, ma2019variational}.
Different quantum algorithms have been applied to industrial fields, for example, SP \cite{eldar2002quantum, low2017optimal},
image processing \cite{yan2017quantum, wang2022review},
quantum architecture search (QAS) \cite{du2022quantum, zhang2022differentiable, fosel2021quantum, ding2022evolutionary, zhang2021neural},
% chemistry \cite{xia2017electronic, streif2019solving},
optimization problem \cite{amaro2022case}, and
scheduling problems \cite{amaro2022case, ikeda2019application, venturelli2015quantum}.
Due to the hardware noises, the performance of
the quantum circuit can be reduced\cite{das2021adapt, zhang2022differentiable}.
Moreover, available qubits are limited,
and cannot solve complex problems without a
training process.

VQA is a leading hybrid quantum algorithm combining quantum and classical computation to solve complex problems \cite{cerezo2021variational}.
In VQA, a quantum circuit with adjustable parameters is optimized using classical
optimizers given a specific objective function.
Combinatorial optimization problems
aim to optimize an objective function within a large configuration
space defined by discrete values while adhering to specific constraints.
VQA, such as VQE and QAOA, is suitable for solving combinatorial optimization problems.

JSSP is one of the combinatorial optimization problems and can be formalized as a
quadratic unconstrained binary
optimization (QUBO) problem. It abstracts the job shop as a model
of a working location containing some available machines and a set of jobs. Each job
includes a sequence of operations. 
% In practice, each machine can be viewed as a signal
% processing module, and each job and its corresponding operations can be considered as signal
% processing tasks and subtasks. 
JSSP has been
solved by VQA using a predefined circuit architecture \cite{amaro2022case}.
However, manually finding a task-specific, hardware-efficient, and noise-resilient quantum circuit is hard. We try to optimize the quantum circuit architecture for solving JSSP.

DQAS, as one of the QAS algorithms, is proposed in \cite{zhang2022differentiable}. It is a gradient-based algorithm and aims to choose a sequence
of unitaries to form a circuit that can optimize a task-specific objective function. DQAS relaxes the parameter determining the circuit structure into a continuous domain, allowing the gradient descent search.
It performs excellently
searching for noise-resilient circuits with error mitigation \cite{du2022quantum, zhang2022differentiable}.
Although QAOA has been successfully tested with DQAS \cite{zhang2022differentiable},
QAOA is often limited by circuit deep and architectures, which may
restrict its performance.
Since VQE views the circuit as an approximator of the ground state and can provide more flexibility, we consider using VQE for our framework JSSP-DQAS to search for a suitable circuit for JSSP.

In this work, we propose a new framework, JSSP-DQAS. It can design
noise-resilient quantum circuit architectures efficiently and automatically.
The searched circuits outperform manually designed circuits and reduce hardware costs with fewer gates by keeping quantum features. Moreover, a shallow circuit
outperforms a complex circuit for a simple task.

\section{Method}
\label{Method}
Algorithm~\ref{alg:DQAS4JSSP} shows an overview of our framework JSSP-DQAS containing
four steps.
The initialization
step defines an operation pool $\mathcal{O}$, a circuit with $p$ placeholders, the
shared architecture parameter $\alpha$ and trainable weights $\theta$ of the parameterized
circuit, an objective function, and a target value. A batch of
circuit architectures is sampled in the training process according to the
probabilistic model $P$. Each architecture calculates the local MSE loss between the
predicted and target values as the objective function.
Then, the global loss $\mathcal{L}$ comes out based on all sampled architectures' local
objective function $L$. They use the gradient-based
optimizer Adam to update $\alpha$, $\theta$, and the probability in $P$ of each
operation candidate.
The top $K$ circuit architectures are determined after the architecture search
procedure. The fine-tuning process starts, which only updates $\theta$, if necessary.
Finally, we get the top $K$ circuit architectures and fine-tuned $\theta$. We use these
designed circuits to solve JSSP and search the circuits again if their performances
are bad.
\begin{algorithm}[t!]
\caption{DQAS for JSSP}\label{alg:DQAS4JSSP}
    \begin{algorithmic}
        \State $\textbf{Step 1: Initialization: }$
        \Indent
        \State Initialize circuit with $p$ placeholders, operation pool $\mathcal{O}$,
               $\alpha$, $\theta$, objective function and target value.
        \EndIndent
        \State $\textbf{Step 2: Super-circuit training: }$
        \Indent
        \While{Architecture search}
        \State Sample a batch of circuit architectures.
        \State Calculate global loss $\mathcal{L}$ via Eq:~\ref{eq:global loss}
        \State Update $\alpha_t$ and  $\theta_t$ via gradients $\nabla\mathcal{L}_\alpha$ %Eq:~\ref{eq:gradient-alpha}\\
        and $\nabla\mathcal{L}_\theta$
        \EndWhile
        \State Circuit parameter tuning
        \EndIndent
        \State $\textbf{Step 3: Get circuits: }$
        \Indent
        \State Take top-K architectures for JSSP evaluation
        \EndIndent
        \State $\textbf{Step 4: Evaluation of architectures: }$
        \Indent
        \State Take the best performing architecture or retrain
        \EndIndent
  \end{algorithmic}
\end{algorithm}

\subsection{Circuit and parameters}
% Initialization defines a circuit, an operation pool $\mathcal{O}$, $\alpha$, $\theta$,
% an objective function and a target value.
The circuit includes encoding, parameterized and measurement blocks. The encoding
block prepares states for the parameterized blocks.
The parameterized block stacked with $p$ placeholders can be seen as a composition
of a sequence of unitaries:
\begin{equation}
    U = \prod ^p _{i=0}u_i(\theta_i) \quad,
\end{equation}
where $\theta_i$ is the trainable parameters of the corresponding unitary in the
sequence. When $u_i$ is a gate with no parameters, $\theta_i$
can be ignored. This work defines operation candidates
and placeholders differently than the original DQAS. Each placeholder $u_i$ covers all
qubits instead of one qubit and
accepts one element from the operation pool containing the
working range.
This way, the number of parameters for each
placeholder depends on the operation type and working range,
and the search space can thus be reduced by controlling the
working range of operation candidates.
The circuit architecture is updated based on the probabilistic model $P(U,\alpha)$.
Each $u_i$ will be placed by operation candidate $o_i \in \mathcal{O}$.

% The measurement block measures the observables and calculates the corresponding
% objective functions.
\subsection{Operation pool}\label{subsection: operation pool}
Operation pool $\mathcal{O}$ in size of $s=|\mathcal{O}|$ is a set of quantum gate
candidates. Each operation in $\mathcal{O}$ contains the type of operation and its
working range.
If there is a circuit with five qubits, an operation pool can be
defined with elements:
\begin{align}
 \mathcal{O}&=\{\underbrace{o_1}_{\textbf{Type}}:\underbrace{[0,1,2,3,4]}_{\textbf{Working range}},
 o_2:[0,1], E:[0,1,2,3,4]\}.
\end{align}
The $o_i$ denotes the type of operation, and it could be any 1-qubit gate or multi-qubit
gate (e.g., \texttt{RZ}, \texttt{U3}, or \texttt{CNOT}). The corresponding array shows
the range in which the operation works. For example, if operation $o_1=$ \texttt{CNOT}
is selected for one placeholder, five \texttt{CNOT} gates will work on all qubits with
ring connections. This work defines two operation pools, op1 and op2.  op1
contains candidates such as \texttt{ry}, \texttt{rz} both working on
$\{[0,1,2,3,4],[0,1,2,3],[1,2,3,4]\}$, \texttt{cz}, \texttt{cnot} working on $[0,1,2,3]$
and identity on all qubits.
op2 contains almost the same candidates but has no \texttt{cz}.

\subsection{Objectives and gradients}
JSSP assigns $J$ jobs
to $M$ machines at specific time slots \cite{amaro2022case}.
Each job is assigned its due
time and processing group. The processing time of each job is often identical.
Each job needs to be processed on
every machine, and the beginning time on the next machine cannot be earlier
than the last. Each machine contains several
idle time slots at the beginning or end. The total time slots of
machine $m$ is \textit{$T_m$ = J + $i_m$}, where $J$ is the number of jobs and $i_m$
is the number of idle time slots of machine $m$.

\begin{figure}[t!]
    \centering
    \includegraphics[width=0.7\linewidth]{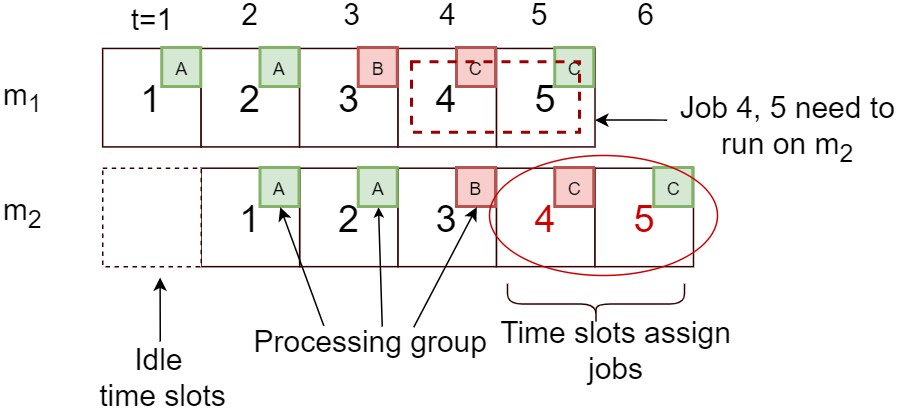}
    \caption{The illustration of a JSSP task.}
    \label{Fig.JSSPTask}
\end{figure}

Fig. \ref{Fig.JSSPTask} illustrates a task of JSSP to be solved in this work, where time slots 1, 5
and 6 on machine 2 are idle, and jobs 4 and 5 need to be assigned on machine 2 in order.
The red circle points out the optimal scheduler of this problem.

A solution of a JSSP consists of two schedules $x\in \mathbb{B}^{N_x} $
and $y\in \mathbb{B}^{N_y} $, where $\mathbb{B} = \{0,1\}$,
${N_x} = \sum_{m=1}^{M}J(J+i_m)$ and ${N_y} = \sum_{m=1}^{M}i_m$. $x$
represents real jobs and $y$ represents dummy jobs that fill idle time slots. A value
$x_{mjt}=1$ represents that job $j$ is assigned to machine
$m$ at time t. The value of a dummy job $y_{mt}=1$ indicates that this
dummy job is assigned to machine $m$ at time $t$.
The quadratic form
$Q:\mathbb{B}^{N_x}\times \mathbb{B}^{N_y}\rightarrow\mathbb{R}$ of JSSP is:
\begin{equation}
    \label{eq:minH}
    (x^*, y^*) = \mathop{\arg\min}_{(x,y)\in \mathbb{B}^{N_x} \times \mathbb{B}^{N_y}}
    Q(x,y) \quad.
\end{equation}

With cost and constraint penalties, the objective function $Q$ becomes:
    \begin{align}
        Q(x,y)&=c(x)+a_1 \sum_{m=1}^{M}\sum_{j=1}^{J}(g_{mj}(x)-1)^2 \\ \nonumber
           & + a_2 \sum_{m=1}^{M}\sum_{t=1}^{T_m}(l_{mt}(x,y)-1)^2 \\
           & + a_3 \sum_{m=1}^{M-1}\sum_{j=1}^{J}q_{mj}(x) +
           a_4 \sum_{m=2}^{M}\sum_{t=1}^{i_m-1}r_{mt}(y) \quad, \nonumber
    \end{align}
where $c(x)$ is the cost and the other terms correspond to constraints for
$1$. job assignment, $2$. time assignment, $3$. process order, and $4$. idle slot.
Details for each cost and constraint term can be found in \cite{amaro2022case}.
$a_1\mbox{-}a_4$ are
coefficients used as weights to make a balance among constraints.

With quantum computing, finding the optimal solution to a QUBO problem is equivalent
to finding the ground state of the Hamiltonian $H$ created by replacing
variables of $Q(x,y)$:
\begin{equation}
    \label{eq:H}
    H = Q(\frac{I-Z_x}{2},\frac{I-Z_y}{2}) \quad,
\end{equation}
where $Z$ indicates PauliZ operators for binary vectors $x$, $y$.

We use CVaR to calculate the cost of JSSP.
To make a simple notation, we use the concatenation of x, y, where Q(z) = Q(x, y).
We sample $k$
bitstrings and calculate the energy $E_k(\theta)$ for each bitstring $b_k$, where
$E_k(\theta)=Q(b_k(\theta))$. These energies are
arranged in ascending order. This sample of energies $\{E_1(\theta),\dots E_k(\theta)\}$
is put into the CVaR estimator to calculate the energy:
\begin{equation}
    C(\theta) =\frac{1}{\left\lceil \alpha K\right\rceil}
    \sum_{k=0}^{\left\lceil \alpha K\right\rceil}E_k(\theta) \quad.
\end{equation}

The global objective function is defined as the total sum of local objectives according
to the architecture distribution model $P(U, \alpha)$
and the energy of the bitstring of the optimal solution $E_{{\rm target}}$:
\begin{align}
    \mathcal{L}=\sum_{U\sim P(U,\alpha)}L(U,\theta)\label{eq:global loss}\quad,
\end{align}
where
\begin{equation}
    \label{eq:JSSP}
    L(U,\theta)=(C(\theta)-E_{{\rm target}})^2 \quad.
\end{equation}

We will iteratively update the parameters by gradient descent, including the circuit
parameter $\theta$ and the architecture parameter $\alpha$. Since $\theta$ is
independent of the architecture distribution, its gradient takes the form:
\begin{align}
    \nabla_{\theta} \mathcal{L}=\sum_{U\sim P(U,\alpha)}\nabla_{\theta} L(U,\theta)\quad.\label{eq:gradient-theta}
\end{align}

In practice, this gradient can be calculated by the parameter shift rules
\cite{crooks2019gradients, wierichs2022general}. The gradient of the architecture parameter matrix
$\alpha$ is related to the architecture distribution and calculated as described in
DQAS \cite{zhang2022differentiable}.

\section{Experiments and results}
\label{Experiments}

\subsection{Experiment setting}
In this chapter, we conduct experiments with five qubits using JSSP-DQAS for the JSSP
task described in Fig. \ref{Fig.JSSPTask} and defined in \cite{amaro2022case}.
Their manually designed circuit, represented
in Fig. \ref{fig: baseline}, is selected as our baseline.
In the experiments, all qubits are initialized with the state $\ket{0}$ and applied
\texttt{rx} gates with a rotation angle of $\pi$ to form the encoding block. There is
only one parameterized block containing four placeholders. For each learning step, we will update all placeholders.

The energies $E$ are scaled to the range $[0,1]$:
\begin{equation}
    e=\frac{E-E_{\rm min}}{E_{\rm max}-E_{\rm min}} \in [0,1] \quad,
\end{equation}
where $E_{\rm min}$ and $E_{\rm max}$ are the minimum and maximum energy.
When $e\approx 0$, we reach the minimal energy and find the
optimal solution.

\subsection{Results and discussion}
\begin{figure}[t!]
    \centering
    \subfloat[Circuit$_{\texttt{op1}}$\label{fig: 1 best structure}]{%
        \includegraphics[width=0.4\linewidth]{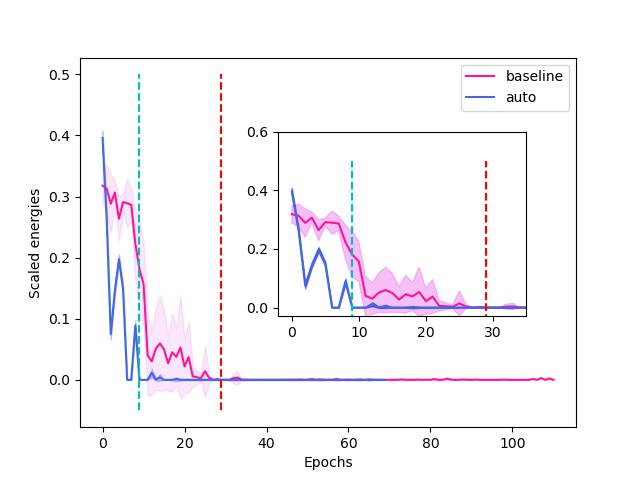}}
    \quad
    \subfloat[Circuit$_{\texttt{op2}}$\label{fig: 2 best structure}]{%
       \includegraphics[width=0.4\linewidth]{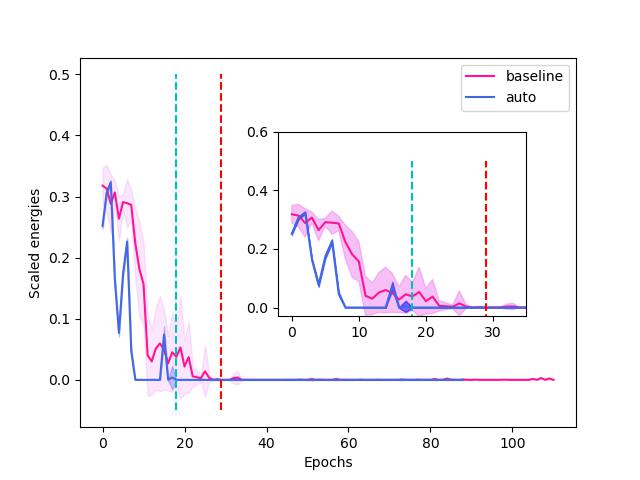}}
    \\
    \subfloat[Baseline\label{fig: baseline}]{
    \includegraphics[width=0.27\linewidth]{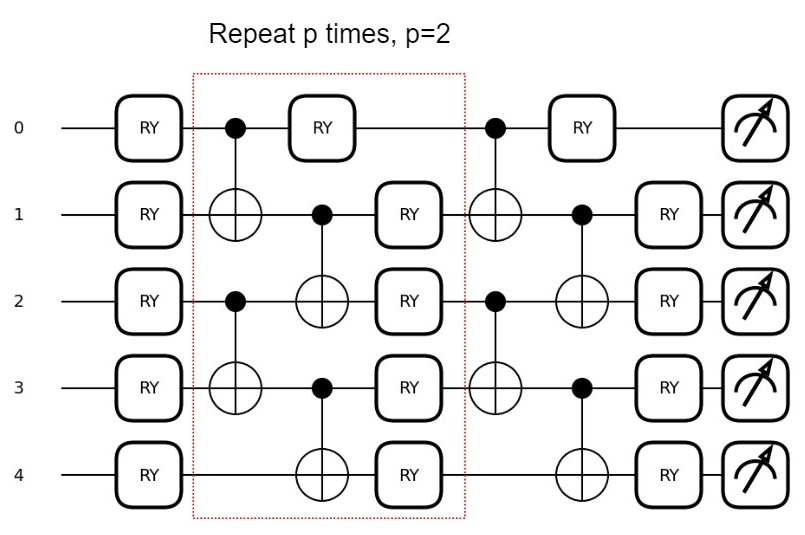}
    }
    \quad
    \subfloat[Circuit$_{\texttt{op1}}$]{
    \includegraphics[width=0.27\linewidth]{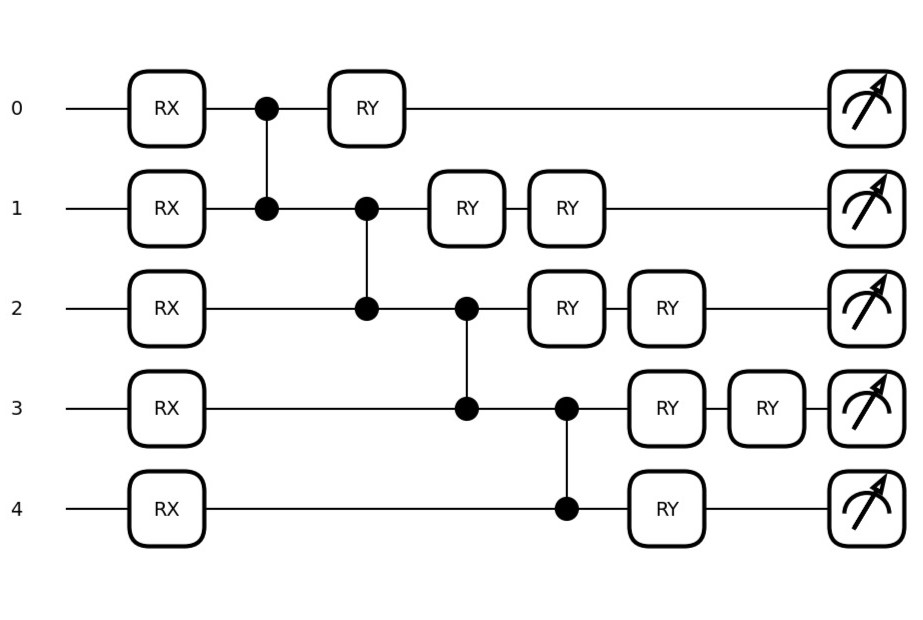}
    \label{fig:Architecture 1}
    }
    \quad
    \subfloat[Circuit$_{\texttt{op2}}$]{
    \includegraphics[width=0.27\linewidth]{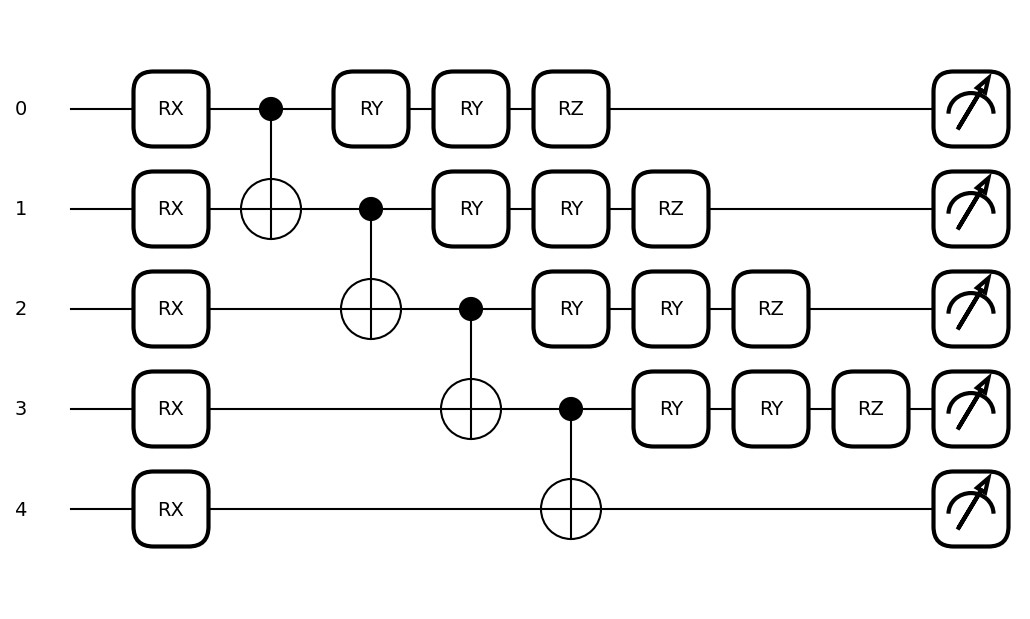}
    \label{fig:Architecture 2}
    }
  \caption{Evaluation of newly found architectures on the simulator
           without noise. The results are averaged over 200 trials with the same
           initial parameters.}
  \label{fig:best structures}
\end{figure}
Fig. \ref{fig:best structures} shows the evaluation of newly found architectures on
the simulator without noise. The automatically designed architectures converge
faster than the baseline to the average solving point (ASP), which means the average
results reach the optimal schedule within 20 epochs. The shadowed standard
deviation areas of the learned circuits are much smaller than the baseline, indicating that
circuits yield consistent outcomes for multiple trials, reflecting the learned
architectures' stability, reliability and reproducibility.

The best architecture Circuit$_{\texttt{op1}}$ derived from op1 in Fig. \ref{fig:Architecture 1}
begins with \texttt{cz} gates. The type of controlled gates is different from that
in the baseline. There are no redundantly parameterized blocks in Circuit$_{\texttt{op1}}$.
Circuit$_{\texttt{op2}}$ derived from op2 in Fig. \ref{fig:Architecture 2} has redundant gates on specific qubits, 
% begins with \texttt{cnot}
% gates and follows with \texttt{ry} and \texttt{rz} gates.
% Generally, manually designed circuits rarely contain redundant gates in one
% parameterized block, and the single-qubit gates are always placed on all qubits. 
which might improve the training performance.

In general, more gates in a circuit,
especially the controlled gates, indicate a higher building cost of the quantum circuit,
while controlled gates provide quantum features by creating entanglements.
Circuit$_{\texttt{op1}}$ and Circuit$_{\texttt{op2}}$ contain fewer gates than the baseline, while
Circuit$_{\texttt{op1}}$ has the fewest. The searched circuits thus have lower
building costs, keeping quantum features.
Fewer parameterized gates and trainable parameters lead to a more
straightforward quantum computation process. This simplification can
improve the overall efficiency of the quantum circuit during training, resulting in
faster convergence. Some noises and errors in a quantum circuit are caused by
decoherence and gate imperfections. Using fewer gates can reduce the
number of operations that cause noises or errors and lead to more reliable and
stable circuits that converge faster.

\begin{figure}[t!]
    \centering
    \subfloat[op1\label{fig: 1 noise}]{%
        \includegraphics[width=0.45\linewidth]{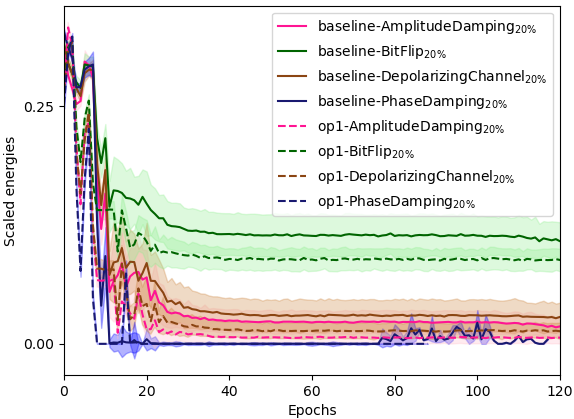}}
    \quad
    \subfloat[op2\label{fig: 2 noise}]{%
       \includegraphics[width=0.45\linewidth]{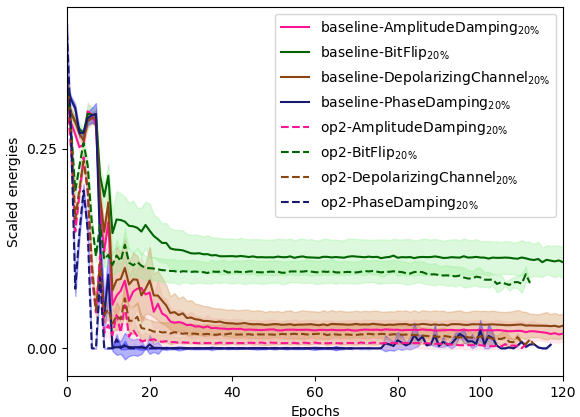}}
  \caption{Evaluation of newly found architectures on noisy simulator.}
  \label{fig:noise evaluation}
\end{figure}

\begin{figure}[t!]
    \centering
    \subfloat[Comparison of op1\label{fig: placeholder}]{%
        \includegraphics[width=0.4\linewidth]{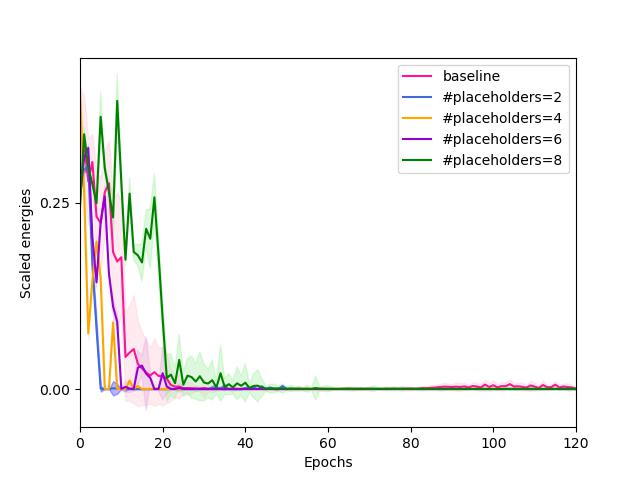}}
    \quad
    \subfloat[Gates and ASP of op1\label{fig: placeholder-gate}]{%
       \includegraphics[width=0.4\linewidth]{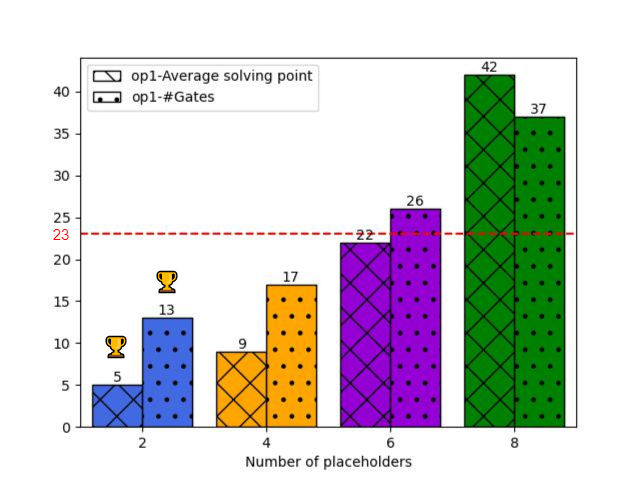}}
    \\
    \subfloat[Comparison of op2\label{fig: layer}]{%
       \includegraphics[width=0.4\linewidth]{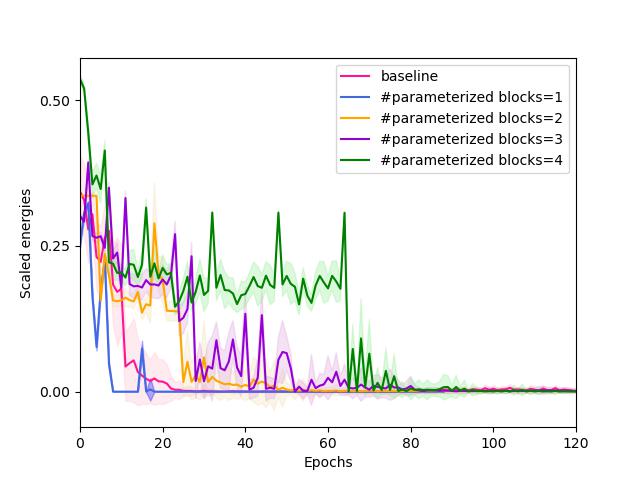}}
   \quad
   \subfloat[Gates and ASP of op2\label{fig: layer-gate}]{%
      \includegraphics[width=0.4\linewidth]{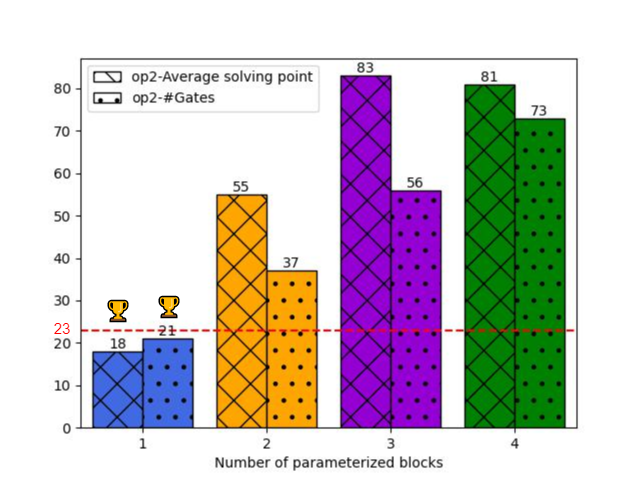}}
  \caption{(a) and (b) illustrate the impact of placeholders when generating
          circuits with op1. (c) and (d) illustrate the impact of parameterized blocks
          when generating circuits with op2.}
  \label{fig:compare}
\end{figure}

In Fig. \ref{fig:noise evaluation}, we study the resistance of auto-generated
circuits to noise. We form noise
models by adding 20\% noise of different types on each end of the qubit.
The shadowed standard deviation areas of designed
architectures are much smaller than the baseline, indicating those learned circuits
yield consistent outcomes. Under BitFlip noise, the baseline and the
searched circuit end with the average results much larger than the minimum energy.
However, the ASP and
the minimum energies found by designed circuits outperform the baseline in different
models.
Although because of the influence of noise, the learned circuits, including the baseline, do not find the accurate minimum energy in some trials, the automatically designed circuits are noise-resilient and more reliable in most cases.

% This tells us these learned circuits reach the accurate minimum
% energy with more trials. However, the baseline does not find the accurate minimum energy in
% some trials because of the influence of noise. Therefore, the automatically designed
% circuits are noise-resilient and more reliable.

In Fig. \ref{fig:compare}, we study the impact of placeholders and parameterized
blocks during generating circuits. The red line indicates the number of gates and
the ASP of the baseline is 23.
The evaluation results show that as the number
of placeholders and parameterized blocks increase, the number of gates in the generated
circuits and the depth of
the circuits increase and converge more slowly when solving a simple JSSP. The deeper the
circuits are, the more parameterized gates need to be
trained, resulting in more complex quantum calculations and slower convergence.

\section{Conclusion and Outlook}
\label{Conclusion and Outlook}
In this work, we propose a framework JSSP-DQAS. The results show that it can design
noise-resilient quantum
circuit architectures efficiently and automatically. The searched
circuits outperform the manually designed circuits. The newly
discovered circuits with fewer
gates reduce the hardware cost by keeping quantum features. Our experiments have
demonstrated the beneficial
impact of DQAS on scheduling problems.
This work uses a shallow circuit and a large learning step for the training
process. However, a better architecture might hide in a deep super-circuit in case of a complex task.

\section*{Acknowledgment}
The project of this workshop paper is based on was supported with funds from the
German Federal Ministry of Education and Research in the funding program
Quantum Reinforcement Learning for industrial Applications (QLindA) and the Federal Ministry for Economic Affairs and Climate Action
in the funding program Quantum-Classical Hybrid Optimization Algorithms for Logistics
and Production Line Management (QCHALLenge)
- under project number 13N15644 and 01MQ22008B. The sole responsibility for the paper’s contents
lies with the authors.
\bibliographystyle{IEEEbib}
\bibliography{strings,refs}

\end{document}